\documentclass[twocolumn]{article}

\usepackage[dvips]{graphicx}
\usepackage{authblk}

\title{Air Mounted Eyepiece: \\
Design Methods for Aerial Optical Functions of \\
Near-Eye and See-Through Display using \\
Transmissive Mirror Device}

\author[1, 2]{Yoichi Ochiai\thanks{wizard@slis.tsukuba.ac.jp, ochyai@pixiedusttech.com}}
\author[1, 2]{Kazuki Otao\thanks{kazuki.otao@pixiedusttech.com}}
\author[1, 2]{Hiroyuki Osone}
\affil[1]{University of Tsukuba}
\affil[2]{Pixie Dust Technologies, Inc.}

\date{}

\begin{document}
\maketitle

\begin{abstract}

We propose a novel method to implement an optical see-through head mounted display which renders real aerial images 
with a wide viewing angle, called an Air Mounted Eyepiece (AME). 
To achieve the AMD design,
we employ an off-the-shelf head mounted display and Transmissive Mirror Device (TMD)
which is usually used in aerial real imaging systems.
In the proposed method,
we replicate the function of the head mounted display (HMD) itself,
which is used in the air by using the TMD and presenting a real image of eyepiece in front of the eye.
Moreover, it can realize a wide viewing angle 3D display
by placing a virtual lens in front of the eye without wearing an HMD.
In addition to enhancing the experience of mixed reality and augmented reality,
our proposed method can be used as a 3D imaging method for use in other applications such as in automobiles and desktop work.
We aim to contribute to the field of human-computer interaction
and the research on eyepiece interfaces
by discussing the advantages and the limitations of this near-eye optical system.

\end{abstract}

\section{Introduction}

\begin{figure*}[h]
 \includegraphics[width=\textwidth]{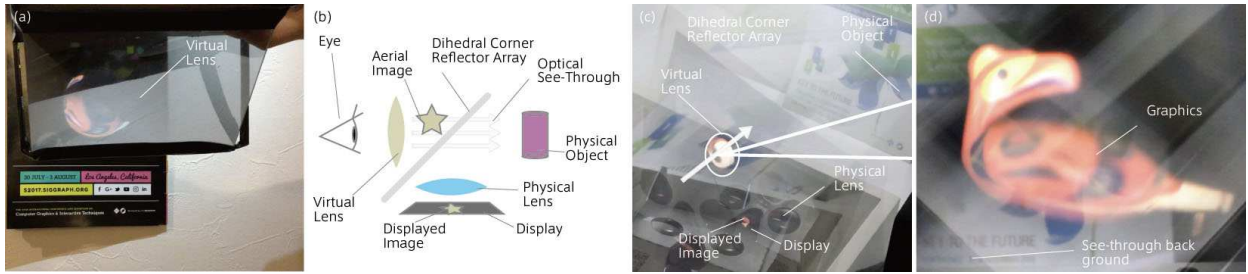}
 \caption{(a) Our prototype. (b) System overview. (c) Air Mounted Eyepiece with virtual lens and physical object. (d) A photo taken of a prototype display with areial graphics and see-though background.}
 \label{fig:teaser}
\end{figure*}

\begin{figure*}[h]
 \includegraphics[width=\textwidth]{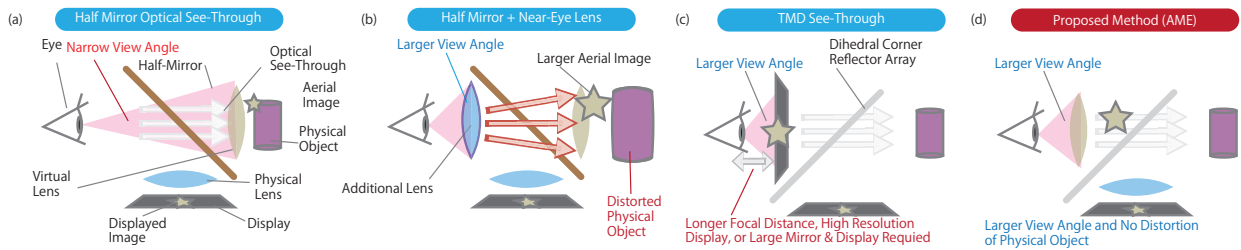}
  \caption{Possible variation of optical see-through aerial imaging by combining a single lens VR and a half mirror or TMD.
  (a) Half mirror optical see-through provides a narrow viewing angle.
  (b) Half mirror and near eye lens provide distorted scenery.
  (c) TMD see-through.
  (d) Proposed method (Air Mounted Eyepiece). }
 \label{fig:relatedwork}
\end{figure*}

In contemporary times, the realization of immersive experiences in Mixed Reality
and Augmented Reality are desirable.
An optical see-through type information presentation method
with a wide viewing angle is required for this purpose.
In recent years, several research studies have proposed employing optical elements
such as transmitted liquid crystal, half mirrors, holographic optical element and waveguide
to achieve these requirements.
However, there are several problems when using these techniques
in an optical see-through system (Figure~\ref{fig:relatedwork}).
When a single lens VR goggle is combined
with a transmissive liquid crystal or half mirror,
the scene transmitted through the lens and the scenery are distorted
owing to the fact that there is a single lens which blocks the lights from the scenery.
In addition, even though the lens
is presented in front of the eye using a half mirror or holographic optical element or waveguide, a wide viewing angle
cannot be achieved because the distance from the eye to the lens becomes longer.
If a convex lens used in optical party, it restricts magnify rate.

In this paper, we introduce a novel method to present ``virtual lenses''
in front of the eyes using a Transmissive Mirror Device (TMD)
with a Micro Dihedral Corner Reflector Array (MDCRA).
A TMD with an MDCRA has been proposed
~\cite{Maekawa:06}~\cite{Miyazaki:13}
and it has been used for various aerial interactions
~\cite{Makino:2016:HMT:2858036.2858481}.
We propose a method to create the functionality of a head-mounted display
in the air by using a TMD to present a real image of eyepiece through single lens VR goggles.
We call this method an Air Mounted Eyepiece (AME).
By looking into the virtual lens in the air,
the user obtains the same experience
as the head mounted display without wearing goggles.
It is also considered useful to provide a see-through and non-wearable
immersive experience with a wide viewing angle from the safety standpoint
for vehicle mounting and medical applications.
This paper introduces the design methods of AME instead of convex mirror
or waveguide or half mirror for building HMD and near-eye displays.

\begin{figure*}
\includegraphics[width=\textwidth]{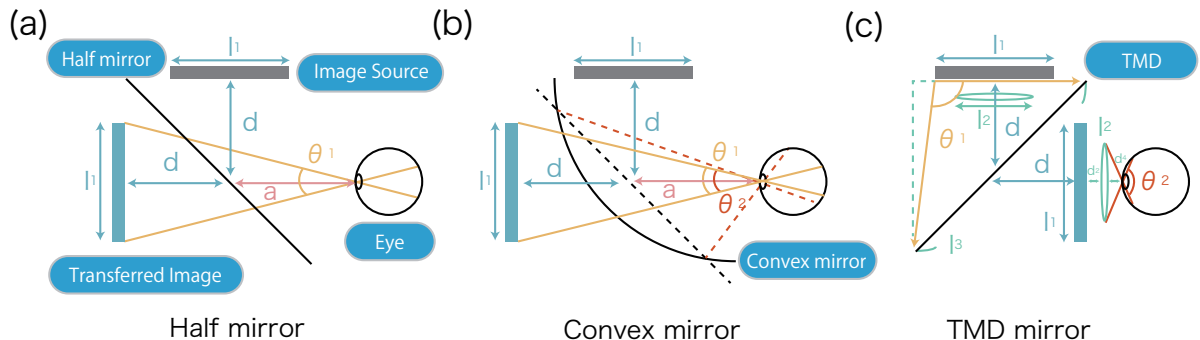}
  \caption{Design method. (a) Half mirror. (b) Convex mirror. (c) TMD mirror.}
\label{fig:desingmethod}
\end{figure*}

\section{Design Method}
\subsubsection*{Aerial Image Formation of the single lens}
We introduce a method to generate the light field of a single lens
as an aerial imaging formulation just close to or inside the eyeball.
By using this method, a transmission type aerial image
with a wider viewing angle than an ordinary transparent HMD can be obtained
as the lens position can be brought closer to the eyeball.
When using an HMD as the light source, the HMD is arranged at the position where the single lens of the HMD is in focus with respect to the TMD eye position, such that they build an optical system that includes the TMD. This allows the aerial imaging of the single lens to be placed in front of the eyeball. By looking into the aerial image presented in front of the eyeball, the same effect as with an HMD can be obtained. Digital transformation is necessary because the image output through the aerial imaging of the single lens is inverted vertically and horizontally.

In Figure.2 we note the parameters in design process of the AME and conventional methods frequently used in see-through HMD design.
Half-mirror based HMD is raised in Figure.~\ref{fig:desingmethod}(a). 
A view angle $\theta_1$ of this setup is 
\begin{equation}
 \frac{l_1}{2(a+d)} = \tan\frac{\theta_1}{2}
\end{equation}
\begin{equation}
 \theta_1 = \tan^{-1}\left(\frac{\frac{l_1}{(a+d)}}{1-\left(\frac{l_1}{(a+d)}\right)^2}\right)
\end{equation}
where $\theta_1$  is size of screen, a is the distance between half-mirror and eyeball, d is distance between half-mirror and screen.
Additional lens are sometimes inserted between screen and half-mirror however it doesn’t change the maximum view angle $\theta_1$. To solve this problem, convex mirror or prisms are used instead of half-mirror. We show in Figure.~\ref{fig:desingmethod}(b).
In this case the view angle $\theta_2$ is
\begin{equation}
\theta_2 = \theta_1 \times a
\end{equation}
where a is the magnification ratio of convex mirror. Note that if a is higher it can not be used as see-through type HMD, because it distorts the see-through view.
Then we introduce our methods by using TMD, in Figure.~\ref{fig:desingmethod}(c), a view angle is
\begin{equation}
if\;\; \theta_1 < \theta_2 : \theta_1 = \tan^{-1}\left(\frac{\frac{l_3}{d_2}}{1-\left(\frac{l_3}{d_2}\right)^2}\right)
\end{equation}
\begin{equation}
if\;\; \theta_1 \geq \theta_2 : \theta_1 = \tan^{-1}\left(\frac{\frac{l_2}{d_4}}{1-\left(\frac{l_2}{d_4}\right)^2}\right)
\end{equation}
where $l_2$ is the size of lens, $l_3$ is the size of TMD, $d_2$ is distance between TMD and screen, $d_4$ is distance between TMD and eyeball. This method can easily be applied to self-designing for HMD in research-prototype, because the $\theta_2$ is shown on the spec sheet of HMD in many cases.  So that prototypers just put HMD close to the HMD on the screen side, it work as see through HMD.

\begin{figure*}
\includegraphics[width=0.9\textwidth]{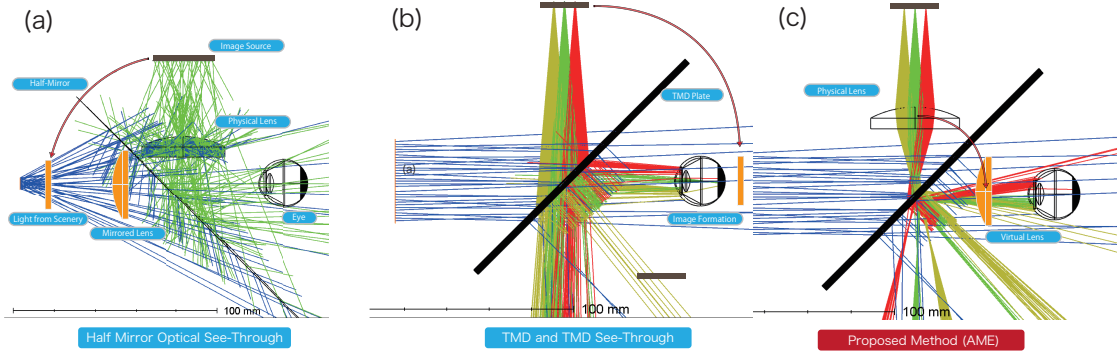}
  \caption{Optical ray tracing. (left) Simulation with half mirror optical see-through configuration. (middle) Simulation with TMD see-though configuration. (right) Simulation with proposed method.}
\label{fig:simulation}
\end{figure*}

\section{Simulation}
Optical ray tracing in a computer simulation 
was performed using Zemax OpticStudio (Figure ~\ref{fig:simulation}).
The TMD structure with a micro dihedral corner reflector array was reproduced with CAD.
The blue line represents an optical ray of light from the scenery.
The red, green, and yellow lines represent the optical ray emitted from the image source.
The half mirror optical see-though display provides a narrow viewing angle.
In our proposed method, the Air Mounted Eyepiece
creates the function of the head mounted display
in the air, thus the virtual lens is placed while in the air.
The relationship between the viewing angle and the change
in the mirror pitch is shown in the following Figure~\ref{fig:tmd-size}.
The higher resolution images are obtained when the mirror pitch is small.

See ~\cite{Miyazaki:13} for more details on the behavior of TMD.

\begin{figure}[t]
 \centering
 \includegraphics[width=3.0in]{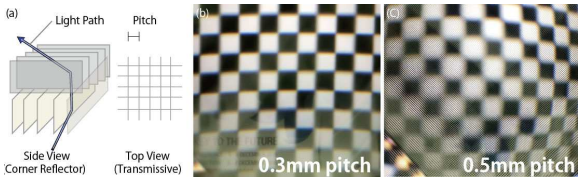}
  \caption{(a) TMD structure which has micro dihedral corner reflector array.
  (b), (c) The image difference of TMD by pitch size.}
 \label{fig:tmd-size}
\end{figure}

\section{Experiment}

\begin{figure}
 \centering
 \includegraphics[width=3.0in]{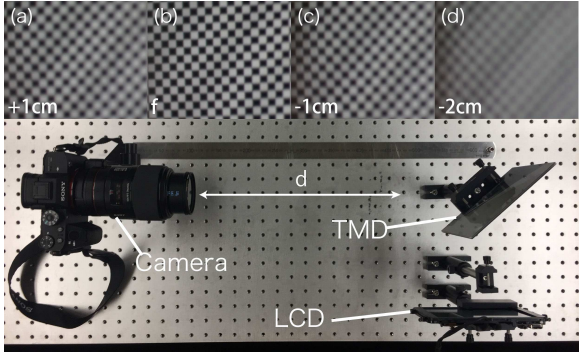}
  \caption{
  Optical setup without eyepiece for the experiment.
  (a) $d = f + 1cm$. 
  (b) $d = f$.
  (c) $d = f - 1cm$. 
  (d) $d = f - 2cm$.}
 \label{fig:no_lens}
\end{figure}

\begin{figure}
 \centering
 \includegraphics[width=3.0in]{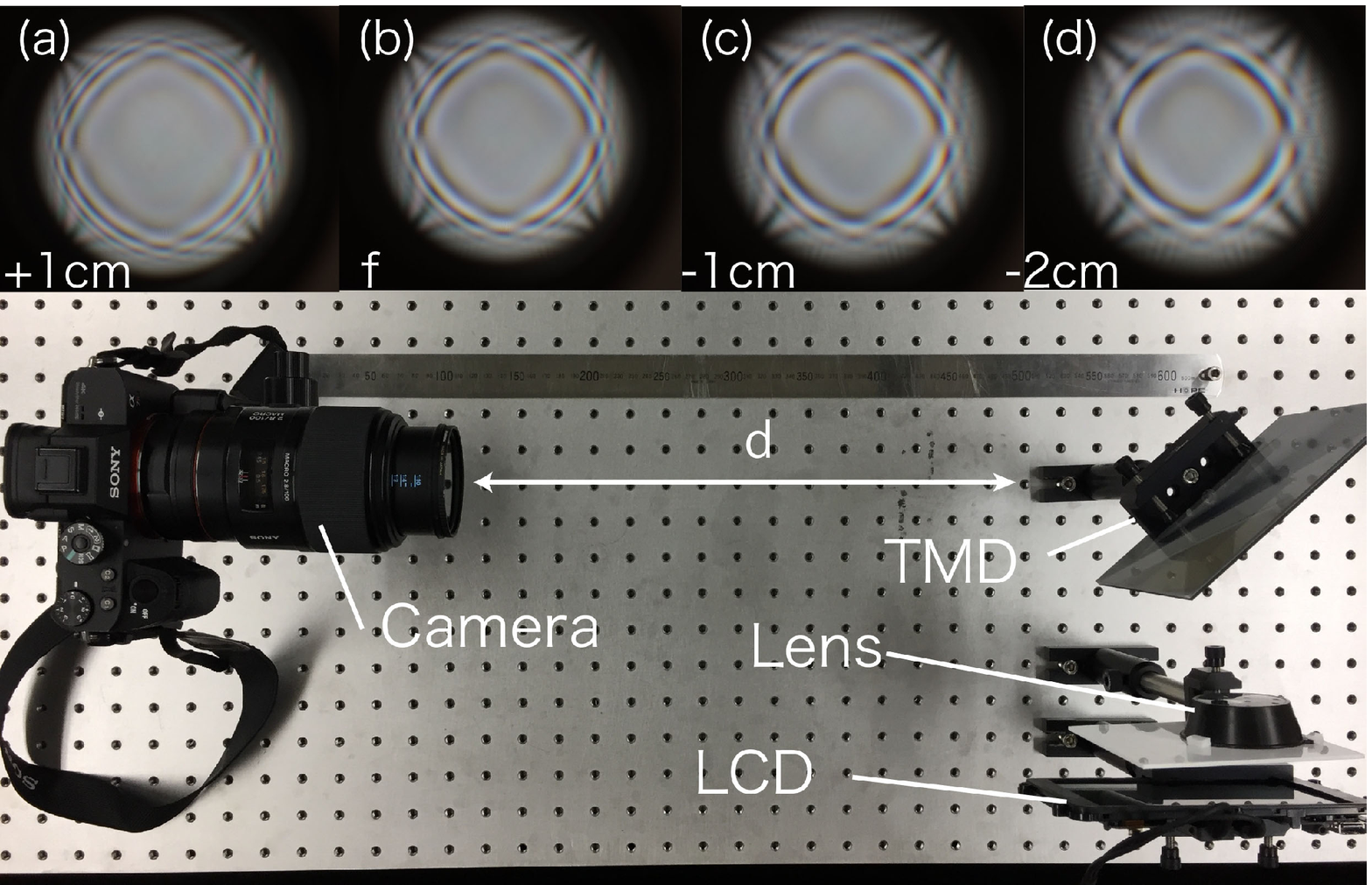}
  \caption{
  Optical setup with eyepiece for the experiment.
  (a) $d = f + 1cm$. 
  (b) $d = f$.
  (c) $d = f - 1cm$. 
  (d) $d = f - 2cm$.}
 \label{fig:lens}
\end{figure}

We conducted an experiment to evaluate AME (Figure~\ref{fig:no_lens} and Figure~\ref{fig:lens}).
The camera (Sony α7R II) was placed at the focal distance.
We moved the camera back and forth, and captured aerial images near the focal length.
Focal length of a camera is $f = 100mm$, ISO sensitivity was $1000$.
LCD and lens (SAL100M28) that are parts of Oculus DK2 were employed.
Pitch size of TMD was $0.5mm$.

\section{Implementation}
\begin{figure*}
 \includegraphics[width=\textwidth]{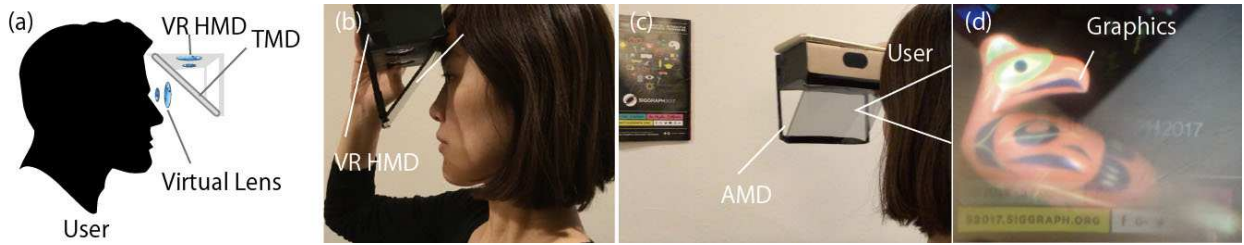}
  \caption{Application of see-though wide view angle HMD.
  (a) System layout.
  (b) Setup.
  (c), (d) Obtained image with aerial image and background scene.}
 \label{fig:app-hmd}
\end{figure*}

\begin{figure*}
 \includegraphics[width=\textwidth]{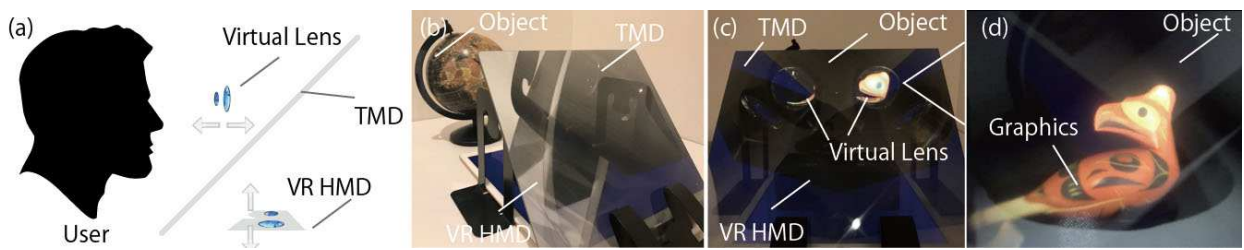}
  \caption{Applicaiton of a non-wearable HMD.
  (a) System layout.
  (b) Setup.
  (c), (d) Obtained image with aerial image and background object.}
 \label{fig:app-amd}
\end{figure*}

We described the design method above with regards to the concept
behind the optical element of the TMD before
and during the aerial imaging of the eye. 
We implemented the application based on that concept.
Figure~\ref{fig:app-hmd} shows the application of the see-through
wide view angle HMD that consists of an off-the-shelf HMD. 
Figure~\ref{fig:app-amd} shows the application of a non-wearable HMD
that enables viewing without wearing goggles.

\subsection*{TMD}
An orthogonal corner mirror cube (a non-commercial product)
was used as the TMD.
The pitch of the mirrors in the mirror cube was 0.5 mm and 0.3 mm,
and the mirror ratio was 1:3.
An aluminum vapor deposition mirror was used. 
The laminated thin plate mirror was cut such that
it was attached at an angle of $45^\circ $ with respect to
the long side of the rectangle.
The following mountings use different mirrors having these pitch values.

\subsection*{Prototype}
\subsubsection*{Aerial Image Formation of the single lens}
A Google cardboard and Oculus DK 2 were used as the HMD in the prototype.
Based on the relationship for the focus in front of the eye,
the maximum viewing angle depends on the viewing angle of each device.
The positional relationships corresponding to the variables described in each viewing angle and resolution are summarized in the Table ~\ref{tab:hmd-performance}.
The image obtained is as shown in the Figure ~\ref{fig:teaser} (d). A polarizing filter was used to prevent the narrowing of the visual field due to the secondary reflected light. The sense of resolution changes depending on the pitch of the mirror. The presumed changes in resolution are summarized in the Figure~\ref{fig:tmd-size}.

\section{Limitation}

\subsection*{Optical Selectivity}
As an optical element that functions by forming an image in the air,
it is only possible to change the light emitting position
and the input position of a light field which possesses the capability
of changing input / output relationship with respect to the light field.
That is, it is difficult to absorb or diffuse light.
In addition, as the resolution is increased,
the mirror functions as a diffraction grating.
With respect to the finer structure,
it is difficult to form imaging and projection systems.
Moreover, because the optical path length is extended,
caution is required when the device is used for applications
such as the ultrashort pulse laser used in Lasik surgery.

\subsection*{TMD Selectivity}
It is known that the corner cube and Nanoink printing TMDs are available on the market.
In this research, a TMD that is easy to obtain and process is prepared and it is easy to perform simulation calculations for this device. 
The same method can be applied to Nanoink print TMDs as well. In fact, there is a high possibility that the resolution of the Nanoink print TMD will be improved.

\section{Conclusion and Futurework}
In this research, we propose a new HMD design that functions as an aerial virtual optical system in front of the eye by using Transmissive Mirror Devices. This is the first challenge: to use a TMD, which is usually used in aerial real imaging systems, for a near-eye display.
We call this method an Air Mounted Eyepiece.
AMD enabled the extension of the conventional HMD to a see-through HMD
with a wide viewing angle. In future research, we will consider gaze tracking. Because the eyeball image is transferable to the HMD side, 
gaze tracking is possible without disturbing the user's line of sight
in wide-view retinal observation. Furthermore, by physically moving the position of the head mount display in accordance with head tracking, an immersive weightless environment can be realized. 
This development will be useful and safe for vehicle mounting and medical applications.

\begin{table}[t]
  \caption{Off-the-shelf HMD performance. Our prototype angle of view depends on each device field of view.} 
  \begin{tabular}{|l|l|l|} \hline
                   & FOV & Resolution      \\ \hline 
  Google cardboard & $90^\circ $   & $1280 \times 800$      \\
                   &               & ($640 \times 800$ per eye) \\ \hline 
  Oculus DK2       & $110^\circ $  & $1920 \times 1080$        \\
                   &               & ($960 \times 1080$ per eye) \\ \hline
  \end{tabular}
  \label{tab:hmd-performance}
\end{table} 

\section{Related Work}

\subsection*{Near-eye See-though Display}
A near-eye see-though display for Augmented Reality
has recently been proposed for various public uses.
In the commercial field,
Microsoft Hololens\footnote{https://www.microsoft.com/en-us/hololens (last accessed October 10, 2017)} and
Meta\footnote{https://www.metavision.com/ (last accessed October 10, 2017)} are available for end users.

Additionally, many optical see-though displays were proposed in previous research.
Displays using half mirrors~\cite{Azuma:1994:ISD:192161.192199} or
free-form prisms~\cite{Hua:14} were proposed.
However, the display with a half mirror distorts
the scene transmitted through the lens
or only allows for a narrow viewing angle,
such that distortion correction has been proposed
~\cite{DBLP:conf/ismar/ItohK15}.
In another method, Waveguide, Retinal scanning,
and Holograpy~\cite{holographic-near-eye-displays-virtual-augmented-reality}
were proposed. However, complex optical configurations or 
high computer performance were required.
Using transmissive liquid crystals enabled optical see-through
and simple HMD configurations
~\cite{conf/ismar/MaimoneF13} employing a stacked transmissive LCD
reproduced the light field
~\cite{Maimone:2014:PDW:2601097.2601141}.
By modulating the point light source, a wide field of view was achieved.
However, there are several problems that block the light from the real scene.

In our proposed method, we employ a TMD as a new optical element
for a near-eye see-through display.
By using a TMD, the real image of a single lens of VR googles is placed in front of the eye,
the function of the head mounted display itself is created in the air.
In contrast to previous methods, the scenery viewed through a TMD is not  distorted
and a wide viewing angle is obtained.

\bibliographystyle{unsrt}
\bibliography{citation}

\end{document}